\pdfoutput=1
\documentclass[aps,11pt,article]{revtex4}
\usepackage{times}
\usepackage{amssymb}
\usepackage{amsmath}
\usepackage{graphicx}
\usepackage{epsfig}
\usepackage{dcolumn}
\usepackage{color}
\usepackage{bm}
\begin{document}
\title{ Nonlinear Ohmic electromagnetic response}

\author{Anwei Zhang}
\email{zhanganwei@shnu.edu.cn}
\affiliation{Department of Physics, Mathematics $\&$ Science College, Shanghai Normal University, No. 100 Guilin Road, Shanghai, 200234 China}
\author{Zheng Cai}
\affiliation{Department of Physics, Mathematics $\&$ Science College, Shanghai Normal University, No. 100 Guilin Road, Shanghai, 200234 China}
\author{C. M. Wang}
\affiliation{Department of Physics, Mathematics $\&$ Science College, Shanghai Normal University, No. 100 Guilin Road, Shanghai, 200234 China}

\begin{abstract}
We systematically investigate nonlinear Ohmic responses in second-harmonic generation and bilinear magnetoelectric effects within the Matsubara Green’s function formalism. The optical nonlinear Ohmic conductivity is shown to consist of a nonlinear Drude-like part and an intrinsic term determined  by the fully symmetrized normalized quantum metric dipole.   Notably, we predict a previously unrecognized intrinsic Ohmic conductivity arising from  band geometry in the bilinear magnetoelectric response, which exhibits transverse behavior similar  to its optical counterpart. Using a two-dimensional Dirac model, we   demonstrate  that this geometrically induced nonlinear Ohmic response  is observable in material  with high Fermi velocity and narrow band gaps. Our  work provides  a  systematic quantum field-theoretic framework  for describing nonlinear Ohmic transport in condensed matter systems.
\end{abstract}

\maketitle

\section{Introduction}
Nonlinear electromagnetic responses have drawn broad interest in recent years, due to their remarkable scientific  value and promising application potential~\cite{du2021nonlinear,10.1093/nsr/nwae334,jiang2025revealing}. Specifically, they offer a unique and powerful perspective for uncovering  the dynamical mechanisms~\cite{sipe2000second,parker2019diagrammatic,holder2020consequences,du2021quantum,zhang2025quantum} and structural symmetries~\cite{gao2014field,sodemann2015quantum,ma2019observation,2023Quantum,zhang2023symmetry} of condensed matter, as well as  for probing   quantum geometric properties~\cite{morimoto2016topological,liu2021intrinsic,bhalla2022resonant,okyay2022second,2024Experimental}, including  the Berry curvature dipole and normalized quantum metric dipole. With steady advances in theory and experiment, nonlinear electromagnetic responses have gradually become a key tool for exploring the essential physical characteristics of periodic crystalline systems and have also laid the groundwork for designing high-efficiency optoelectronic components  and quantum functional devices~\cite{ma2021topology}.

While previous research  on nonlinear electromagnetic responses focuses primarily    on nonlinear Hall effects, recent experiments  have revealed   the existence of nonlinear Ohmic responses in condensed-matter systems  under  zero magnetic field~\cite{2022Graphene,min2024colossal,2023Quantum,PhysRevLett.132.046303}. These experiments further imply the presence of a quantum metric dipole and demonstrate that the nonlinear Ohmic conductivity does not originate from the Berry curvature dipole. Thus, the nonlinear Ohmic response provides a direct means to distinguish between different microscopic origins. Moreover, it offers a physical platform for diode-like transport, radio-frequency energy harvesting, and nonlinear memory devices, thus broadening the research horizon of nonlinear transport phenomena~\cite{suarez2025nonlinear}.
Theoretically, conventional   semiclassical equations governing the motion of electron wave packets fail to capture such nonlinear Ohmic behaviors~\cite{gao2014field,10.1093/nsr/nwae334}. Although refined approaches, including  perturbation-corrected semiclassical formalisms~\cite{kaplan2024unification} and the density matrix method~\cite{das2023intrinsic},  can predict the emergence of nonlinear Ohmic responses and relate them to the quantum metric dipole, their results exhibit considerable inconsistencies with each other. Furthermore, these approaches  are  restricted to electric-field-driven  scenarios and do not account for magnetic field effects. 
Given the absence of a unified, systematic quantum theoretical description, comprehensive investigations into nonlinear Ohmic responses are urgently needed to establish a complete quantum theory.

In this paper, we systematically investigate the nonlinear Ohmic behavior in both the second-harmonic generation (SHG) optical response and the second-order bilinear magnetoelectric response. We adopt the Matsubara Green’s function formalism rooted in  quantum field theory—a  fully  quantum framework to describe intricate many-body phenomena in condensed matter systems.  Within this theoretical framework, we prove that, for both types of response, the nonlinear Ohmic conductivity receives no contribution at linear order in the relaxation time, leaving the conductivity purely Hall. For the optical response, the  Ohmic conductivity consists of  a nonlinear Drude-like component and an intrinsic contribution  governed by the  fully symmetrized normalized quantum metric dipole, which can be transverse.  In the bilinear magnetoelectric response, we identify a new type of   intrinsic nonlinear Ohmic  contribution originating from  band geometry, whose  analytical form is distinct from yet analogous to its optical counterpart.  Beyond Ohmic transport, we also provide the corresponding Hall responses for both cases. To clarify the physical significance, we use  a two-dimensional Dirac model and demonstrate that the Ohmic conductivity in the bilinear magnetoelectric response can be large enough  to be observable in materials with high Fermi velocity and narrow band gaps.

\section{Setup}
We start by considering the second-order response to the electric field product. The conductivity for SHG optical response can be derived using Dyson's equation within the Matsubara Green's function formalism~\cite{zhang2025quantum}, or alternatively through the path integral method combined with diagrammatic techniques~\cite{parker2019diagrammatic,jiang2025revealing}
\begin{eqnarray}\label{q1}
\sigma_{\mu\nu\gamma
}&=&\frac{e^3}{(\omega_0+i \tau^{-1})^2\beta V}\sum_{\mathbf{k},i\omega_n}\mathrm{Tr} \bigg[ \frac{1}{2}\partial_{\mu}\partial_{\nu}\hat{v}_{\gamma}(\mathbf{k})G(\mathbf{k},i\omega_n)+\partial_{\mu}\hat{v}_{\gamma}(\mathbf{k})G(\mathbf{k},i\omega_n)\hat{v}_{\nu}(\mathbf{k})G(\mathbf{k},i\omega_n-i\omega_0)\nonumber\\&&+\frac{1}{2}\hat{v}_{\mu}(\mathbf{k})G(\mathbf{k},i\omega_n)\partial_{\nu}\hat{v}_{\gamma}(\mathbf{k})G(\mathbf{k},i\omega_n-2i\omega_0)\nonumber\\&&+\hat{v}_{\mu}G(\mathbf{k},i\omega_n)\hat{v}_{\nu}G(\mathbf{k},i\omega_n-i\omega_0)\hat{v}_{\gamma}G(\mathbf{k},i\omega_n-2i\omega_0)\bigg]+(\nu\leftrightarrow \gamma),
\end{eqnarray}
Here, $-e$ denotes the charge of the electron, $\omega_0$ represents the frequency of the incident photon, $\beta = 1/k_B T$ is  the inverse temperature, $V$ stands for the volume in a three-dimensional system or the area in a two-dimensional system, $\partial_{\mu} \equiv \partial_{k_\mu}$ indicates  the derivative with respect to the crystal momentum component $k_\mu$, and $\hat{v}_{\gamma}$ is  the velocity operator in the $\gamma$ direction. We set $\hbar=1$  in this paper.
$G(\mathbf{k}, i\omega_n) = \sum_a |u_a(\mathbf{k})\rangle\langle u_a(\mathbf{k})| / (i\omega_n +\mu - \varepsilon_a(\mathbf{k}))$  is the Matsubara Green's function, where $|u_a(\mathbf{k})\rangle$ denotes   the periodic part of the Bloch wave function, $\omega_n = (2n+1)\pi/\beta$ refers to   the fermionic Matsubara frequency, $\mu$ is the chemical potential, and $\varepsilon_a(\mathbf{k})$  is the energy dispersion of band $a$. Repeated indices are summed over. In $\sigma_{\mu\nu\gamma}$, the first index indicates the direction of the induced current, while the last two indices correspond to  the polarization directions of the incident electric fields. Considering  the adiabatic switching of the external fields~\cite{passos2018nonlinear}, the phenomenological relaxation time    $\tau$  is included in the global coefficient, and the analytical continuation of the Matsubara  Green's function now becomes $i\omega_0 \rightarrow \omega_0+i \tau^{-1}$. 

It is well known that for the linear response  $\sigma_{\mu\nu}$,  the dissipationless Hall conductivity $\sigma^{H}_{\mu\nu}$ is antisymmetric in the indices  $\mu$ and $\nu$, given by 
$\sigma^{H}_{\mu\nu}=(\sigma_{\mu\nu}-\sigma_{\nu\mu})/2$, while the dissipative Ohmic conductivity 
$\sigma^{O}_{\mu\nu}=\sigma_{\mu\nu}-\sigma^{H}_{\mu\nu}=(\sigma_{\mu\nu}+\sigma_{\nu\mu})/2$ is symmetric. As a generalization to the SHG response, the Ohmic conductivity can be obtained by symmetrizing over all indices as \cite{tsirkin2022separation}
\begin{equation}\label{q2}
 \sigma^{O}_{\mu\nu\gamma}=\frac{1}{3}(\sigma_{\mu\nu\gamma}+\sigma_{\nu\mu\gamma}+\sigma_{\gamma\nu\mu}),
\end{equation}
where we have used the intrinsic permutation symmetry of  
the conductivity with respect to the last two indices.
Consequently, the Hall conductivity is given by~\cite{he2026second}
\begin{eqnarray}\label{q3}
\sigma^{H}_{\mu\nu\gamma}=\sigma_{\mu\nu\gamma}-\sigma^{O}_{\mu\nu\gamma}=\frac{1}{3}(2\sigma_{\mu\nu\gamma}-\sigma_{\nu\mu\gamma}-\sigma_{\gamma\nu\mu}).
\end{eqnarray}
It can be verified that the Hall conductivity satisfies 
$\sigma^{H}_{\mu\nu\gamma}=-\sigma^{H}_{\nu\mu\gamma}-\sigma^{H}_{\gamma\nu\mu}$ upon exchanging $\mu$ with $\nu$ and $\gamma$, while preserving the  intrinsic  permutation symmetry of the last two indices: $\sigma^{H}_{\mu\nu\gamma}=\sigma^{H}_{\mu\gamma\nu}$. It should be noted that in this study, the terms Ohmic response and Hall response refer to dissipative and non-dissipative responses, respectively, rather than to longitudinal and transverse responses. In fact, Hall and transverse responses are not generally equivalent, and the Ohmic response can itself be transverse~\cite{suarez2025nonlinear}.

\section{Ohmic conductivity in  SHG optical response}
Next, we  consider the above SHG Ohmic conductivity in the DC limit, i.e., $\omega_0 \rightarrow 0$. We expand the  conductivity in terms of $\tau^{-1}$  under this DC limit. 
To begin with, we analyze the $\tau^{2}$-dependent term in the expansion:
\begin{eqnarray}\label{q4}
\sigma^{(O,2)}_{\mu\nu\gamma}&=&\frac{-\tau^2e^3}{3\beta V}\sum_{\mathbf{k},i\omega_n}\mathrm{Tr} \bigg[ \partial_{\mu}\partial_{\nu}\hat{v}_{\gamma}(\mathbf{k})G(\mathbf{k},i\omega_n)+\partial_{\mu}\hat{v}_{\gamma}(\mathbf{k})G(\mathbf{k},i\omega_n)\hat{v}_{\nu}(\mathbf{k})G(\mathbf{k},i\omega_n)\nonumber\\&&+\partial_{\mu}\hat{v}_{\nu}(\mathbf{k})G(\mathbf{k},i\omega_n)\hat{v}_{\gamma}(\mathbf{k})G(\mathbf{k},i\omega_n)+\hat{v}_{\mu}(\mathbf{k})G(\mathbf{k},i\omega_n)\partial_{\nu}\hat{v}_{\gamma}(\mathbf{k})G(\mathbf{k},i\omega_n)\nonumber\\&&+\hat{v}_{\mu}G(\mathbf{k},i\omega_n)\hat{v}_{\nu}G(\mathbf{k},i\omega_n)\hat{v}_{\gamma}G(\mathbf{k},i\omega_n)+\hat{v}_{\mu}G(\mathbf{k},i\omega_n)\hat{v}_{\gamma}G(\mathbf{k},i\omega_n)\hat{v}_{\nu}G(\mathbf{k},i\omega_n)\bigg]\nonumber\\&&+(\mu\leftrightarrow \nu)+(\mu\leftrightarrow \gamma)\nonumber\\
&=&\frac{-\tau^2e^3}{V}\sum_{\mathbf{k},a}f_a(\mathbf{k})\partial_{\mu}\partial_{\nu}\partial_{\gamma}\varepsilon_a(\mathbf{k}).
\end{eqnarray}
Here,  $f_a(\mathbf{k})=1/\big(e^{\beta(\varepsilon_a(\mathbf{k})-\mu)}+1\big)$ denotes the Fermi-Dirac distribution function. This term gives rise to the third derivative of the energy dispersion, corresponding to a nonlinear Drude-like contribution to the Ohmic conductivity. 

We now proceed to the term linear in $\tau$:
\begin{eqnarray}\label{q5}
\sigma^{(O,1)}_{\mu\nu\gamma}&=&\frac{\tau e^3}{3i\beta V}\sum_{\mathbf{k},i\omega_n}\mathrm{Tr} \bigg[\partial_{\mu}\hat{v}_{\gamma}(\mathbf{k})G^{'}(\mathbf{k},i\omega_n)\hat{v}_{\nu}(\mathbf{k})G(\mathbf{k},i\omega_n)+\partial_{\mu}\hat{v}_{\nu}(\mathbf{k})G^{'}(\mathbf{k},i\omega_n)\hat{v}_{\gamma}(\mathbf{k})G(\mathbf{k},i\omega_n)\nonumber\\&&-2\hat{v}_{\mu}(\mathbf{k})G(\mathbf{k},i\omega_n)\partial_{\nu}\hat{v}_{\gamma}(\mathbf{k})G^{'}(\mathbf{k},i\omega_n)+\hat{v}_{\mu}G^{'}(\mathbf{k},i\omega_n)\hat{v}_{\nu}G(\mathbf{k},i\omega_n)\hat{v}_{\gamma}G(\mathbf{k},i\omega_n)\nonumber\\&&-\hat{v}_{\mu}G(\mathbf{k},i\omega_n)\hat{v}_{\nu}G(\mathbf{k},i\omega_n)\hat{v}_{\gamma}G^{'}(\mathbf{k},i\omega_n)+\hat{v}_{\mu}G^{'}(\mathbf{k},i\omega_n)\hat{v}_{\gamma}G(\mathbf{k},i\omega_n)\hat{v}_{\nu}G(\mathbf{k},i\omega_n)\nonumber\\&&-\hat{v}_{\mu}G(\mathbf{k},i\omega_n)\hat{v}_{\gamma}G(\mathbf{k},i\omega_n)\hat{v}_{\nu}G^{'}(\mathbf{k},i\omega_n)\bigg]+(\mu\leftrightarrow \nu)+(\mu\leftrightarrow \gamma),
\end{eqnarray}
which can be found to vanish. Here, $G^{'}(\mathbf{k},i\omega_n)$ denotes the derivative of the Green's function with respect to the imaginary frequency. In deriving  this result,  we have have made use of  both the translational invariance in frequency and the cyclic invariance of the trace. This analysis demonstrates  that, at this order, no Ohmic conductivity emerges, and  the SHG conductivity is purely Hall-like, expressed as: 
$\frac{\tau e^3}{V}\sum_{\mathbf{k},a\ne b}f_a(\mathbf{k})\big(\partial_{\gamma}F^{ab}_{\mu\nu}+\partial_{\nu}F^{ab}_{\mu\gamma}\big)$, a form that respects  time-reversal symmetry~\cite{sodemann2015quantum,kaplan2024unification,zhang2025quantum}. Here,  $F^{ab}_{\mu\nu}=-2\mathrm{lm}\langle \partial_{\mu}u_a(\mathbf{k})\vert  u_b(\mathbf{k})\rangle \langle u_b(\mathbf{k})\vert  \partial_{\nu} u_a(\mathbf{k})\rangle$ denotes  the Berry curvature. The above results are consistent with  previous experimental findings~\cite{2022Graphene,2023Quantum,PhysRevLett.132.046303}  that nonlinear Ohmic conductivity does not arise from the Berry curvature dipole $\partial_{\gamma}F^{ab}_{\mu\nu}$.

Finally, we examine the $\tau$-independent term. 
From Eqs.~(\ref{q1}) and ~(\ref{q2}), we have the Ohmic conductivity in the DC limit:
\begin{eqnarray}\label{q6}
\sigma^{O}_{\mu\nu\gamma}&=&\frac{e^3}{3(i\tau^{-1})^2V}\sum_{\mathbf{k},a,b,c}\bigg[\frac{f_{ab}}{\varepsilon_{ab}-i \tau^{-1}}\langle u_b(\mathbf{k})|\partial_{\mu}\hat{v}_{\gamma}(\mathbf{k})|u_a(\mathbf{k})\rangle\langle u_a(\mathbf{k})|\hat{v}_{\nu}(\mathbf{k})|u_b(\mathbf{k})\rangle \nonumber\\&&+\frac{f_{ab}}{2(\varepsilon_{ab}-2i \tau^{-1})}\langle u_b(\mathbf{k})|\hat{v}_{\mu}(\mathbf{k})|u_a(\mathbf{k})\rangle\langle u_a(\mathbf{k})|\partial_{\nu}\hat{v}_{\gamma}(\mathbf{k})|u_b(\mathbf{k})\rangle\nonumber\\&&+\langle u_c(\mathbf{k})|\hat{v}_{\mu}(\mathbf{k})|u_a(\mathbf{k})\rangle\langle u_a(\mathbf{k})|\hat{v}_{\nu}(\mathbf{k})|u_b(\mathbf{k})\rangle
\langle u_b(\mathbf{k})|\hat{v}_{\gamma}(\mathbf{k})|u_c(\mathbf{k})\rangle\nonumber\\&&\times \frac{1}{\varepsilon_{ab}-i \tau^{-1}}\bigg(\frac{f_{ac}}{\varepsilon_{ac}-2i \tau^{-1}}-\frac{f_{bc}}{\varepsilon_{bc}-i \tau^{-1}}\bigg)+(\nu\leftrightarrow \gamma)
\bigg]+(\mu\leftrightarrow \nu)+(\mu\leftrightarrow \gamma),
\end{eqnarray}
where $f_{ab}\equiv f_a(\mathbf{k})-f_b(\mathbf{k})$.  Here we omit the first term in   Eq.~(\ref{q1}), which only contributes to  $\tau^2$-dependent contributions. For a two-band system, the indices $a,b,c$ can refer only to the valence and conduction bands. If these indices correspond to the same band, Eq.~(\ref{q6}) vanishes identically. Therefore, the  non-vanishing index combinations for $a,b,c$ are $a,b,a;a,b,b$; and $a,a,b$. By expanding  Eq.~(\ref{q6}) in powers of  $\tau^{-1}$ and exchanging the indices $a \leftrightarrow b$ in terms containing $f_b(\mathbf{k})$, we obtain the $\tau$-independent term:
\begin{eqnarray}\label{q7}
\sigma^{(O,0)}_{\mu\nu\gamma}&=&\frac{e^3}{V}\sum_{\mathbf{k},a\ne b}f_a(\mathbf{k})\mathrm{Re}\bigg[\frac{4}{\varepsilon^3_{ab}}\langle u_a(\mathbf{k})|\hat{v}_{\mu}(\mathbf{k})|u_b(\mathbf{k})\rangle\langle u_b(\mathbf{k})|\partial_{\nu}\hat{v}_{\gamma}(\mathbf{k})|u_a(\mathbf{k})\rangle\nonumber\\&&+\frac{4}{\varepsilon^3_{ab}}\langle u_a(\mathbf{k})|\hat{v}_{\nu}(\mathbf{k})|u_b(\mathbf{k})\rangle\langle u_b(\mathbf{k})|\partial_{\mu}\hat{v}_{\gamma}(\mathbf{k})|u_a(\mathbf{k})\rangle\nonumber\\&&+\frac{4}{\varepsilon^3_{ab}}\langle u_a(\mathbf{k})|\hat{v}_{\gamma}(\mathbf{k})|u_b(\mathbf{k})\rangle\langle u_b(\mathbf{k})|\partial_{\mu}\hat{v}_{\nu}(\mathbf{k})|u_a(\mathbf{k})\rangle\nonumber\\&&-\frac{10}{\varepsilon^4_{ab}}(v_{a\mu}-v_{b\mu})\langle u_a(\mathbf{k})|\hat{v}_{\nu}(\mathbf{k})|u_b(\mathbf{k})\rangle
\langle u_b(\mathbf{k})|\hat{v}_{\gamma}(\mathbf{k})|u_a(\mathbf{k})\rangle\nonumber\\&&
-\frac{10}{\varepsilon^4_{ab}}(v_{a\gamma}-v_{b\gamma})\langle u_a(\mathbf{k})|\hat{v}_{\mu}(\mathbf{k})|u_b(\mathbf{k})\rangle\langle u_b(\mathbf{k})|\hat{v}_{\nu}(\mathbf{k})|u_a(\mathbf{k})\rangle\nonumber\\&&
-\frac{10}{\varepsilon^4_{ab}}(v_{a\nu}-v_{b\nu})
\langle u_a(\mathbf{k})|\hat{v}_{\mu}(\mathbf{k})|u_b(\mathbf{k})\rangle\langle u_b(\mathbf{k})|\hat{v}_{\gamma}(\mathbf{k})|u_a(\mathbf{k})\rangle\bigg]
\nonumber\\&=&
\frac{2e^3}{V}\sum_{\mathbf{k},a\ne b}f_a(\mathbf{k})\bigg[\partial_{\nu}\bigg(\frac{g_{\mu\gamma}}{\varepsilon_{ab}}\bigg)+\partial_{\gamma}\bigg(\frac{g_{\mu\nu}}{\varepsilon_{ab}}\bigg)+\partial_{\mu}\bigg(\frac{g_{\nu\gamma}}{\varepsilon_{ab}}\bigg)\bigg],
\end{eqnarray}
where $\varepsilon_{ab}\equiv \varepsilon_a(\mathbf{k})-\varepsilon_b(\mathbf{k})$ denotes
the energy difference between two bands at a given momentum, $v_{a\mu}= \langle u_a(\mathbf{k})|\hat{v}_{\mu}(\mathbf{k})|u_a(\mathbf{k})\rangle=\partial_{\mu}\varepsilon_{a}(\mathbf{k})$  is the group velocity along the $\mu$ direction, and $g_{\mu\gamma}=\mathrm{Re}\langle u_a(\mathbf{k})|\hat{v}_{\mu}(\mathbf{k})|u_b(\mathbf{k})\rangle\langle u_b(\mathbf{k})|\hat{v}_{\gamma}(\mathbf{k})|u_a(\mathbf{k})\rangle/\varepsilon^2_{ab}$ represents  the quantum metric~\cite{provost1980riemannian,zhang2022revealing}.  The quantities $2g_{\mu\nu}/\varepsilon_{ab}$ and $\partial_{\gamma}(2g_{\mu\nu}/\varepsilon_{ab})$  correspond to the band-energy normalized quantum metric and the normalized quantum metric dipole, respectively~\cite{wang2023quantum}.
The intrinsic Hall conductivity is obtained by subtracting this Ohmic contribution from the total conductivity at this order, yielding $\frac{2e^3}{V}\sum_{\mathbf{k},a\ne b}f_a(\mathbf{k})\bigg[\partial_{\nu}\bigg(\frac{g_{\mu\gamma}}{\varepsilon_{ab}}\bigg)+\partial_{\gamma}\bigg(\frac{g_{\mu\nu}}{\varepsilon_{ab}}\bigg)-2\partial_{\mu}\bigg(\frac{g_{\nu\gamma}}{\varepsilon_{ab}}\bigg)\bigg]$,
a form that explicitly requires the breaking of time-reversal symmetry~\cite{gao2014field}.
Eq.~(\ref{q7}) identifies the fully symmetrized normalized quantum metric dipole as the origin of intrinsic nonlinear Ohmic conductivity. This result  aligns with previous experimental observations~\cite{2023Quantum} upon  including  the  quantum metric dipole contribution, while  preserving the  requirement of Ohmic conductivity, i.e., symmetrization across all indices.
It should be noted that the skew scattering~\cite{2022Graphene} and side-jump mechanisms~\cite{PhysRevLett.132.046303}  fall outside the scope of this paper, and that the above results are similar to those derived using the density matrix method~\cite{das2023intrinsic}.

\section{Ohmic response in second-order  magnetoelectric  effect}
In this section, we  extend our analysis  to the second-order bilinear magnetoelectric response and identify  a new type of nonlinear Ohmic response. We now take into account the wavevector $\mathbf{q}_0$ of the external electromagnetic fields. From Ref.~\cite{zhang2025quantum}, the coefficient of the second-order Ohmic response  $\langle \hat{j}_{\mu}(2\mathbf{q}_0,2i\omega_0) \rangle$ associated with the vector potential product 
 $A_{\nu}(\mathbf{q}_0,\omega_0)A_{\gamma}(\mathbf{q}_0,\omega_0)$ in the SHG regime can be expressed as  
\begin{eqnarray}\label{r1}
\Pi^{O}_{\mu\nu\gamma
}&=&\frac{-e^3}{3\beta V}\sum_{\mathbf{k},i\omega_n}\mathrm{Tr} \bigg[ \frac{1}{2}\partial_{\mu}\partial_{\nu}\hat{v}_{\gamma}(\mathbf{k})G(\mathbf{k},i\omega_n)\nonumber\\&&+\partial_{\mu}\hat{v}_{\gamma}(\mathbf{k})G(\mathbf{k}+\mathbf{q}_0/2,i\omega_n+i\omega_0/2)\hat{v}_{\nu}(\mathbf{k})G(\mathbf{k}-\mathbf{q}_0/2,i\omega_n-i\omega_0/2)\nonumber\\&&+\frac{1}{2}\hat{v}_{\mu}(\mathbf{k})G(\mathbf{k}+\mathbf{q}_0,i\omega_n+i\omega_0)\partial_{\nu}\hat{v}_{\gamma}(\mathbf{k})G(\mathbf{k}-\mathbf{q}_0,i\omega_n-i\omega_0)\nonumber\\&&+\hat{v}_{\mu}G(\mathbf{k}+\mathbf{q}_0,i\omega_n+i\omega_0)\hat{v}_{\nu}G(\mathbf{k},i\omega_n)\hat{v}_{\gamma}G(\mathbf{k}-\mathbf{q}_0,i\omega_n-i\omega_0)+(\nu\leftrightarrow \gamma)\bigg]\nonumber\\&&+(\mu\leftrightarrow \nu)+(\mu\leftrightarrow \gamma).
\end{eqnarray}
Here, analogous to the symmetrization procedure employed for the Ohmic conductivity in Eq.~(\ref{q2}), we symmetrize the expression over all indices. We also exploit the translational invariance in both wavevector and frequency, and omit terms involving ${q_{0}}^2_\eta\partial^2_{\eta}\hat{v}_{\gamma}$.
The above expression is thus exact up to first order in $\mathbf{q}_0$, which is sufficiently accurate for describing the bilinear magnetoelectric response.

Upon performing the analytic continuation $i\omega_0 \rightarrow \omega_0+i \tau^{-1}$,  converting the vector potential to the electric field via  $\mathbf{A}(\mathbf{q}_0,\omega_0)=\mathbf{E}(\mathbf{q}_0,\omega_0)/i(\omega_0+i\tau^{-1})$, expanding the coefficient to first order in  $\mathbf{q}_0$ to give the magnetic field 
$\mathbf{B}(\mathbf{q}_0,\omega_0)=i\mathbf{q}_0\times \mathbf{A}(\mathbf{q}_0,\omega_0)$, and taking the DC limit $\omega_0 \rightarrow 0$, the first order  contribution in $\tau$ — corresponding to the zeroth order term in $\tau$ for the above coefficient in Eq.~(\ref{r1}) — can be found to be zero. This result demonstrates  that the bilinear magnetoelectric effect exhibits no Ohmic response at first order in  $\tau$, and only a Hall response survives~\cite{zhang2023nonlinear}.

We now turn to  the $\tau$-independent term. From Eq.~(\ref{r1}), we have the Ohmic conductivity $\sigma_{\mu\nu}=\langle \hat{j}_{\mu}(2\mathbf{q}_0,2\omega_0) \rangle/E_{\nu}(\mathbf{q}_0,\omega_0)=\Pi^{O}_{\mu\nu\gamma}A_{\gamma}(\mathbf{q}_0,\omega_0)
/i(\omega_0+i\tau^{-1})$ in the DC limit:
\begin{eqnarray}\label{r2}
\sigma^{O}_{\mu\nu}&=&\frac{\tau e^3}{3V}\sum_{\mathbf{k},a, b,c}\bigg\{\bigg[\langle u_b(\mathbf{k}-\mathbf{q}_0/2)|\partial_{\mu}\hat{v}_{\gamma}(\mathbf{k})|u_a(\mathbf{k}+\mathbf{q}_0/2)\rangle\langle u_a(\mathbf{k}+\mathbf{q}_0/2)|\hat{v}_{\nu}(\mathbf{k})|u_b(\mathbf{k}-\mathbf{q}_0/2)\rangle \nonumber\\&&\times\frac{f_a(\mathbf{k}+\mathbf{q}_0/2)-f_b(\mathbf{k}-\mathbf{q}_0/2)}{\varepsilon_a(\mathbf{k}+\mathbf{q}_0/2)-\varepsilon_b(\mathbf{k}-\mathbf{q}_0/2)-i \tau^{-1}}+\frac{f_a(\mathbf{k}+\mathbf{q}_0)-f_b(\mathbf{k}-\mathbf{q}_0)}{2(\varepsilon_a(\mathbf{k}+\mathbf{q}_0)-\varepsilon_b(\mathbf{k}-\mathbf{q}_0)-2i \tau^{-1})}\nonumber\\&&\times\langle u_b(\mathbf{k}-\mathbf{q}_0)|\hat{v}_{\mu}(\mathbf{k})|u_a(\mathbf{k}+\mathbf{q}_0)\rangle\langle u_a(\mathbf{k}+\mathbf{q}_0)|\partial_{\nu}\hat{v}_{\gamma}(\mathbf{k})|u_b(\mathbf{k}-\mathbf{q}_0)\rangle\nonumber\\&&+\frac{1}{\varepsilon_a(\mathbf{k}+\mathbf{q}_0)-\varepsilon_b(\mathbf{k})-i \tau^{-1}}\bigg(\frac{f_a(\mathbf{k}+\mathbf{q}_0)-f_c(\mathbf{k}-\mathbf{q}_0)}{\varepsilon_a(\mathbf{k}+\mathbf{q}_0)-\varepsilon_c(\mathbf{k}-\mathbf{q}_0)-2i \tau^{-1}}-\frac{f_b(\mathbf{k})-f_c(\mathbf{k}-\mathbf{q}_0)}{\varepsilon_b(\mathbf{k})-\varepsilon_c(\mathbf{k}-\mathbf{q}_0)-i \tau^{-1}}\bigg)\nonumber\\&&\times\langle u_c(\mathbf{k}-\mathbf{q}_0)|\hat{v}_{\mu}(\mathbf{k})|u_a(\mathbf{k}+\mathbf{q}_0)\rangle\langle u_a(\mathbf{k}+\mathbf{q}_0)|\hat{v}_{\nu}(\mathbf{k})|u_b(\mathbf{k})\rangle
\langle u_b(\mathbf{k})|\hat{v}_{\gamma}(\mathbf{k})|u_c(\mathbf{k}-\mathbf{q}_0)\rangle+(\nu\leftrightarrow\gamma)\bigg]\nonumber\\&&+(\mu\leftrightarrow \nu)+(\mu\leftrightarrow \gamma)\bigg\}A_{\gamma}(\mathbf{q}_0,\omega_0).
\end{eqnarray}
Here we omit the first term in Eq.~(\ref{r1}), as it is independent of the wavevector and thus gives no contribution to the effect.
For a two-band system, Eq. (\ref{r2}) can be decomposed into intraband and interband contributions. The intraband component, corresponding to transitions within the same band but at distinct momentum space points, takes the form:
\begin{eqnarray}\label{r3}
\sigma^{O,intra}_{\mu\nu}&=&\frac{\tau e^3}{3V}\sum_{\mathbf{k},a\ne b}\bigg\{\bigg[\frac{1}{\varepsilon_{a}(\mathbf{k}+\mathbf{q}_0)-\varepsilon_{b}(\mathbf{k})-i \tau^{-1}}\frac{f_a(\mathbf{k}+\mathbf{q}_0)-f_a(\mathbf{k}-\mathbf{q}_0)}{\varepsilon_{a}(\mathbf{k}+\mathbf{q}_0)-\varepsilon_{a}(\mathbf{k}-\mathbf{q}_0)-2i \tau^{-1}}\nonumber\\&&\times\langle u_a(\mathbf{k}-\mathbf{q}_0)|\hat{v}_{\mu}(\mathbf{k})|u_a(\mathbf{k}+\mathbf{q}_0)\rangle\langle u_a(\mathbf{k}+\mathbf{q}_0)|\hat{v}_{\nu}(\mathbf{k})|u_b(\mathbf{k})\rangle
\langle u_b(\mathbf{k})|\hat{v}_{\gamma}(\mathbf{k})|u_a(\mathbf{k}-\mathbf{q}_0)\rangle\nonumber\\&&+\frac{-1}{\varepsilon_{b}(\mathbf{k}+\mathbf{q}_0)-\varepsilon_{a}(\mathbf{k})-i \tau^{-1}}\frac{f_a(\mathbf{k})-f_a(\mathbf{k}-\mathbf{q}_0)}{\varepsilon_a(\mathbf{k})-\varepsilon_a(\mathbf{k}-\mathbf{q}_0)-i \tau^{-1}}\nonumber\\&&\times\langle u_a(\mathbf{k}-\mathbf{q}_0)|\hat{v}_{\mu}(\mathbf{k})|u_b(\mathbf{k}+\mathbf{q}_0)\rangle\langle u_b(\mathbf{k}+\mathbf{q}_0)|\hat{v}_{\nu}(\mathbf{k})|u_a(\mathbf{k})\rangle
\langle u_a(\mathbf{k})|\hat{v}_{\gamma}(\mathbf{k})|u_a(\mathbf{k}-\mathbf{q}_0)\rangle\nonumber\\&&+\frac{1}{\varepsilon_{a}(\mathbf{k})-\varepsilon_{b}(\mathbf{k}-\mathbf{q}_0)-i \tau^{-1}}\frac{f_a(\mathbf{k}+\mathbf{q}_0)-f_a(\mathbf{k})}{\varepsilon_{a}(\mathbf{k}+\mathbf{q}_0)-\varepsilon_a(\mathbf{k})-i \tau^{-1}}\nonumber\\&&\times\langle u_b(\mathbf{k}-\mathbf{q}_0)|\hat{v}_{\mu}(\mathbf{k})|u_a(\mathbf{k}+\mathbf{q}_0)\rangle\langle u_a(\mathbf{k}+\mathbf{q}_0)|\hat{v}_{\nu}(\mathbf{k})|u_a(\mathbf{k})\rangle
\langle u_a(\mathbf{k})|\hat{v}_{\gamma}(\mathbf{k})|u_b(\mathbf{k}-\mathbf{q}_0)\rangle\nonumber\\&&+(\nu\leftrightarrow\gamma)\bigg]+(\mu\leftrightarrow \nu)+(\mu\leftrightarrow \gamma)\bigg\}A_{\gamma}(\mathbf{q}_0,\omega_0).
\end{eqnarray}
Upon differentiating   Eq.~(\ref{r3}) with respect to ${q_{0}}_\eta$ and expanding it to zeroth order in $\tau$,  one  obtains the $\tau$-independent term
\begin{equation}\label{r4}
\sigma^{(O,0),intra}_{\mu\nu}=\frac{-2ie^3}{V}\sum_{\mathbf{k},a\ne b} f^{'}_a(\mathbf{k}) \bigg(\frac{g_{\mu\gamma}}{\varepsilon_{ab}} v_{a\nu}+\frac{g_{\mu\nu}}{\varepsilon_{ab}} v_{a\gamma}+\frac{g_{\nu\gamma}}{\varepsilon_{ab}} v_{a\mu}\bigg){q_0}_{\eta}A_{\gamma}(\mathbf{q}_0,\omega_0).
\end{equation}
Here, $f^{'}_a(\mathbf{k})\equiv\partial_{\eta}f_a(\mathbf{k})$.
The interband part corresponds to transitions between different bands and reads:
\begin{eqnarray}\label{r5}
\sigma^{O,inter}_{\mu\nu}&=&\frac{\tau e^3}{3V}\sum_{\mathbf{k},a\ne b}\bigg\{\bigg[\langle u_b(\mathbf{k}-\mathbf{q}_0/2)|\partial_{\mu}\hat{v}_{\gamma}(\mathbf{k})|u_a(\mathbf{k}+\mathbf{q}_0/2)\rangle\langle u_a(\mathbf{k}+\mathbf{q}_0/2)|\hat{v}_{\nu}(\mathbf{k})|u_b(\mathbf{k}-\mathbf{q}_0/2)\rangle \nonumber\\&&\times\frac{f_a(\mathbf{k}+\mathbf{q}_0/2)-f_b(\mathbf{k}-\mathbf{q}_0/2)}{\varepsilon_a(\mathbf{k}+\mathbf{q}_0/2)-\varepsilon_b(\mathbf{k}-\mathbf{q}_0/2)-i \tau^{-1}}+\frac{f_a(\mathbf{k}+\mathbf{q}_0)-f_b(\mathbf{k}-\mathbf{q}_0)}{2(\varepsilon_a(\mathbf{k}+\mathbf{q}_0)-\varepsilon_b(\mathbf{k}-\mathbf{q}_0)-2i \tau^{-1})}\nonumber\\&&\times\langle u_b(\mathbf{k}-\mathbf{q}_0)|\hat{v}_{\mu}(\mathbf{k})|u_a(\mathbf{k}+\mathbf{q}_0)\rangle\langle u_a(\mathbf{k}+\mathbf{q}_0)|\partial_{\nu}\hat{v}_{\gamma}(\mathbf{k})|u_b(\mathbf{k}-\mathbf{q}_0)\rangle\nonumber\\&&+\frac{-1}{\varepsilon_{a}(\mathbf{k}+\mathbf{q}_0)-\varepsilon_{b}(\mathbf{k})-i \tau^{-1}}\frac{f_a(\mathbf{k}-\mathbf{q}_0)-f_b(\mathbf{k})}{\varepsilon_{a}(\mathbf{k}-\mathbf{q}_0)-\varepsilon_{b}(\mathbf{k})+i \tau^{-1}}\nonumber\\&&\times\langle u_a(\mathbf{k}-\mathbf{q}_0)|\hat{v}_{\mu}(\mathbf{k})|u_a(\mathbf{k}+\mathbf{q}_0)\rangle\langle u_a(\mathbf{k}+\mathbf{q}_0)|\hat{v}_{\nu}(\mathbf{k})|u_b(\mathbf{k})\rangle
\langle u_b(\mathbf{k})|\hat{v}_{\gamma}(\mathbf{k})|u_a(\mathbf{k}-\mathbf{q}_0)\rangle\nonumber\\&&+\frac{1}{\varepsilon_{a}(\mathbf{k}+\mathbf{q}_0)-\varepsilon_{b}(\mathbf{k})-i \tau^{-1}}\frac{f_a(\mathbf{k}+\mathbf{q}_0)-f_b(\mathbf{k}-\mathbf{q}_0)}{\varepsilon_a(\mathbf{k}+\mathbf{q}_0)-\varepsilon_b(\mathbf{k}-\mathbf{q}_0)-2i \tau^{-1}}\nonumber\\&&\times\langle u_b(\mathbf{k}-\mathbf{q}_0)|\hat{v}_{\mu}(\mathbf{k})|u_a(\mathbf{k}+\mathbf{q}_0)\rangle\langle u_a(\mathbf{k}+\mathbf{q}_0)|\hat{v}_{\nu}(\mathbf{k})|u_b(\mathbf{k})\rangle
\langle u_b(\mathbf{k})|\hat{v}_{\gamma}(\mathbf{k})|u_b(\mathbf{k}-\mathbf{q}_0)\rangle\nonumber\\&&+\frac{-1}{\varepsilon_{a}(\mathbf{k}+\mathbf{q}_0)-\varepsilon_{b}(\mathbf{k}-\mathbf{q}_0)-2i \tau^{-1}}\frac{f_a(\mathbf{k}+\mathbf{q}_0)-f_b(\mathbf{k}-\mathbf{q}_0)}{\varepsilon_{a}(\mathbf{k})-\varepsilon_{b}(\mathbf{k}-\mathbf{q}_0)-i \tau^{-1}}\nonumber\\&&\times\langle u_b(\mathbf{k}-\mathbf{q}_0)|\hat{v}_{\mu}(\mathbf{k})|u_a(\mathbf{k}+\mathbf{q}_0)\rangle\langle u_a(\mathbf{k}+\mathbf{q}_0)|\hat{v}_{\nu}(\mathbf{k})|u_a(\mathbf{k})\rangle
\langle u_a(\mathbf{k})|\hat{v}_{\gamma}(\mathbf{k})|u_b(\mathbf{k}-\mathbf{q}_0)\rangle\nonumber\\&&+(\nu\leftrightarrow\gamma)\bigg]+(\mu\leftrightarrow \nu)+(\mu\leftrightarrow \gamma)\bigg\}A_{\gamma}(\mathbf{q}_0,\omega_0).
\end{eqnarray}
For simplicity, we here consider a two-band system without an overall energy shift, such that the quantity $\partial_{\alpha}(\varepsilon_{a}+\varepsilon_{b})=v_{a\alpha}+v_{b\alpha}$ vanishes. Differentiating   Eq.~(\ref{r5}) with respect to ${q_{0}}_\eta$ and expanding to zeroth order in $\tau$ yields
\begin{eqnarray}\label{r6}
\sigma^{(O,0),inter}_{\mu\nu}&=&\frac{ie^3}{V}\sum_{\mathbf{k},a\ne b} f^{'}_a(\mathbf{k})\bigg\{\frac{4}{\varepsilon_{ab}}\big[g_{\nu\gamma} (v_{b\mu}-v_{a\mu})+g_{\mu\nu}(v_{b\gamma}-v_{a\gamma})+g_{\mu\gamma} (v_{b\nu}-v_{a\nu})\big]\nonumber\\&&+\frac{2}{\varepsilon^2_{ab}}\mathrm{Re}\big[\langle u_a(\mathbf{k})|\hat{v}_{\mu}(\mathbf{k})|u_b(\mathbf{k})\rangle\langle u_b(\mathbf{k})|\partial_{\nu}\hat{v}_{\gamma}(\mathbf{k})|u_a(\mathbf{k})\rangle\nonumber\\&&+\langle u_a(\mathbf{k})|\hat{v}_{\nu}(\mathbf{k})|u_b(\mathbf{k})\rangle\langle u_b(\mathbf{k})|\partial_{\mu}\hat{v}_{\gamma}(\mathbf{k})|u_a(\mathbf{k})\rangle\nonumber\\&&+\langle u_a(\mathbf{k})|\hat{v}_{\gamma}(\mathbf{k})|u_b(\mathbf{k})\rangle\langle u_b(\mathbf{k})|\partial_{\mu}\hat{v}_{\nu}(\mathbf{k})|u_a(\mathbf{k})\rangle\big]\bigg\}{q_0}_{\eta}A_{\gamma}(\mathbf{q}_0,\omega_0)\nonumber\\&=&\frac{ie^3}{V}\sum_{\mathbf{k},a\ne b} f^{'}_a(\mathbf{k})(\partial_{\nu}g_{\mu\gamma}+\partial_{\gamma}g_{\mu\nu}+\partial_{\mu}g_{\nu\gamma}){q_0}_{\eta}A_{\gamma}(\mathbf{q}_0,\omega_0).
\end{eqnarray}
Combining the intraband and interband contributions, we obtain the intrinsic Ohmic conductivity which is independent of $\tau$:
\begin{eqnarray}\label{r7}
\sigma^{(O,0)}_{\mu\nu}&=&\frac{ie^3}{V}\sum_{\mathbf{k},a\ne b} f^{'}_a(\mathbf{k}) \bigg[(\partial_{\nu}g_{\mu\gamma}+\partial_{\gamma}g_{\mu\nu}+\partial_{\mu}g_{\nu\gamma})-2\bigg(\frac{g_{\mu\gamma}}{\varepsilon_{ab}} v_{a\nu}+\frac{g_{\mu\nu}}{\varepsilon_{ab}} v_{a\gamma}+\frac{g_{\nu\gamma}}{\varepsilon_{ab}} v_{a\mu}\bigg)\bigg]{q_0}_{\eta}A_{\gamma}(\mathbf{q}_0,\omega_0)\nonumber\\&=&\frac{ie^3}{V}\sum_{\mathbf{k},a\ne b} f^{'}_a(\mathbf{k}) \bigg\{\bigg[\partial_{\nu}\bigg(\frac{g_{\mu\gamma}}{\varepsilon_{ab}}\bigg)+\partial_{\gamma}\bigg(\frac{g_{\mu\nu}}{\varepsilon_{ab}}\bigg)+\partial_{\mu}\bigg(\frac{g_{\nu\gamma}}{\varepsilon_{ab}}\bigg)\bigg]\varepsilon_{ab}\bigg\}{q_0}_{\eta}A_{\gamma}(\mathbf{q}_0,\omega_0).
\end{eqnarray}
Here, we have used the relation $v_{a\alpha}+v_{b\alpha}=0$. Following a similar procedure, the intrinsic Hall conductivity for this second-order bilinear magnetoelectric  response is found  to be
\begin{eqnarray}\label{r8}
\sigma^{(H,0)}_{\mu\nu}&=&\frac{ie^3}{V}\sum_{\mathbf{k},a\ne b} f^{'}_a(\mathbf{k}) \bigg\{\frac{2}{\varepsilon_{ab}}\big[g_{\mu\nu}(v_{b\gamma}-v_{a\gamma})+g_{\mu\gamma} (v_{b\nu}-v_{a\nu})-2g_{\nu\gamma} (v_{b\mu}-v_{a\mu})\big]\nonumber\\&&+\frac{1}{\varepsilon^2_{ab}}\mathrm{Re}\big[2\langle u_a(\mathbf{k})|\hat{v}_{\mu}(\mathbf{k})|u_b(\mathbf{k})\rangle\langle u_b(\mathbf{k})|\partial_{\nu}\hat{v}_{\gamma}(\mathbf{k})|u_a(\mathbf{k})\rangle\nonumber\\&&-\langle u_a(\mathbf{k})|\hat{v}_{\nu}(\mathbf{k})|u_b(\mathbf{k})\rangle\langle u_b(\mathbf{k})|\partial_{\mu}\hat{v}_{\gamma}(\mathbf{k})|u_a(\mathbf{k})\rangle\nonumber\\&&-\langle u_a(\mathbf{k})|\hat{v}_{\gamma}(\mathbf{k})|u_b(\mathbf{k})\rangle\langle u_b(\mathbf{k})|\partial_{\mu}\hat{v}_{\nu}(\mathbf{k})|u_a(\mathbf{k})\rangle\big]\bigg\}{q_0}_{\eta}A_{\gamma}(\mathbf{q}_0,\omega_0)\nonumber\\&=&\frac{ie^3}{V}\sum_{\mathbf{k},a\ne b} f^{'}_a(\mathbf{k})\bigg \{\bigg[\partial_{\nu}\bigg(\frac{g_{\mu\gamma}}{\varepsilon_{ab}}\bigg)+\partial_{\gamma}\bigg(\frac{g_{\mu\nu}}{\varepsilon_{ab}}\bigg)-2\partial_{\mu}\bigg(\frac{g_{\nu\gamma}}{\varepsilon_{ab}}\bigg)\bigg]\varepsilon_{ab}\bigg\}{q_0}_{\eta}A_{\gamma}(\mathbf{q}_0,\omega_0).
\end{eqnarray}
It can be seen that the conductivity structures given in Eqs.~(\ref{r7}) and (\ref{r8}) resemble those of SHG optical responses. Specifically, both Ohmic conductivities include the fully symmetrized term $\partial_{\nu}\bigg(\frac{g_{\mu\gamma}}{\varepsilon_{ab}}\bigg)+\partial_{\gamma}\bigg(\frac{g_{\mu\nu}}{\varepsilon_{ab}}\bigg)+\partial_{\mu}\bigg(\frac{g_{\nu\gamma}}{\varepsilon_{ab}}\bigg)$, whereas the Hall conductivities contain the term $\partial_{\nu}\bigg(\frac{g_{\mu\gamma}}{\varepsilon_{ab}}\bigg)+\partial_{\gamma}\bigg(\frac{g_{\mu\nu}}{\varepsilon_{ab}}\bigg)-2\partial_{\mu}\bigg(\frac{g_{\nu\gamma}}{\varepsilon_{ab}}\bigg)$ which vanishes identically when $\mu=\nu=\gamma$.
Nevertheless, owing to the even symmetry of the quantum metric and energy dispersion under time reversal,  optical conductivities necessarily require time-reversal symmetry breaking, whereas bilinear magnetoelectric conductivities can emerge in time-reversal-invariant systems.

\begin{figure}
\centering
\includegraphics[width=0.5
\textwidth]{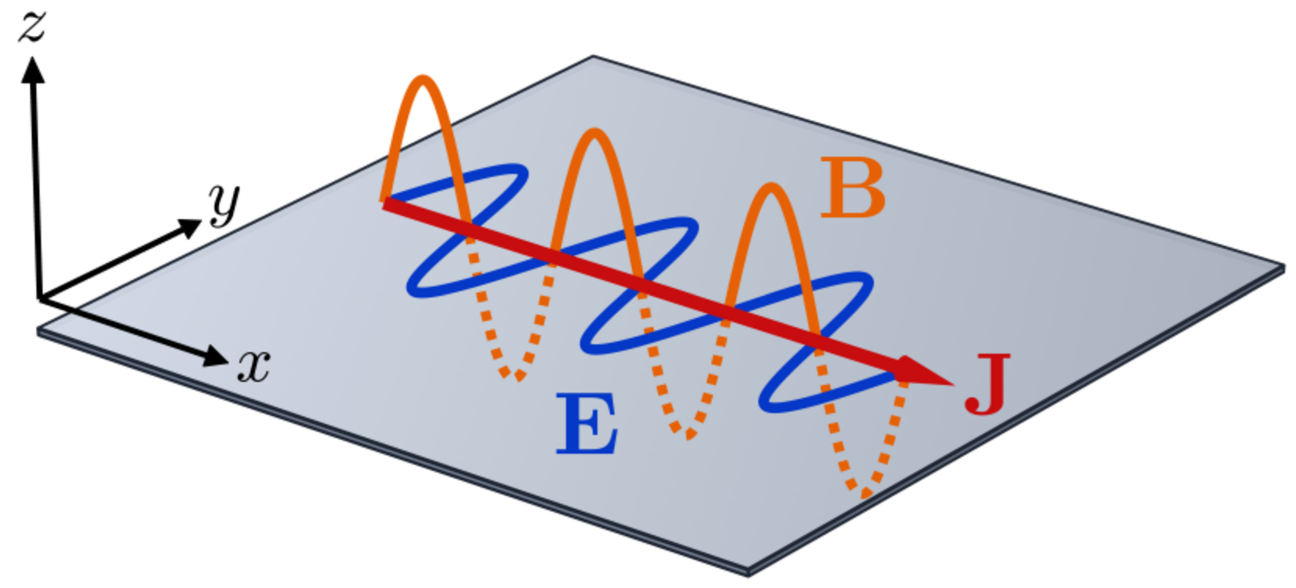}
\caption{Schematic of the considered system. The electric field $\mathbf{E}$ and magnetic field $\mathbf{B}$ are polarized along the $y$- and $z$-directions, respectively. The induced Hall current $\mathbf{J}$ is perpendicular to the electromagnetic fields which propagate along the same direction as the current.
}\label{figure.1}
\end{figure}

\section{Application to two-dimensional Dirac model}
\begin{figure}
\centering
\includegraphics[width=0.5
\textwidth]{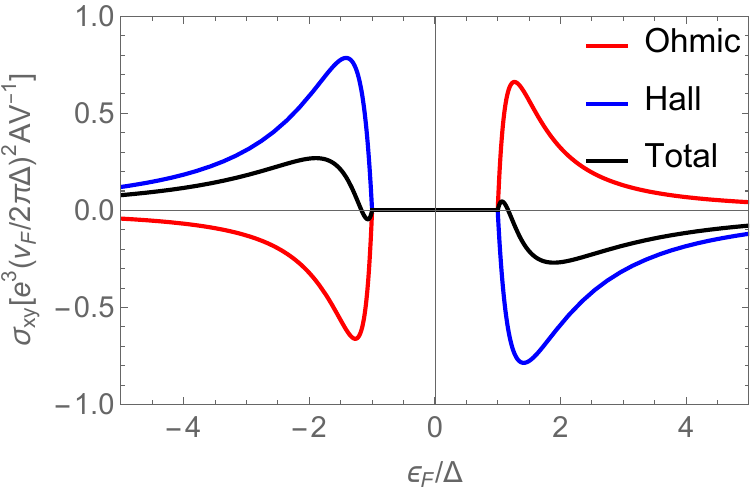}
\caption{The Ohmic and Hall conductivities as a function of $\epsilon_F/\Delta$. Here, the magnetic filed is set to 1 T.
}\label{figure.2}
\end{figure}

In what follows, we apply the above predicted intrinsic Ohmic conductivity in the bilinear magnetoelectric response to  a massive Dirac model at the $sK$ valley.  The model is  described by the Hamiltonian $H= v_Fk_x\sigma_x+v_Fk_y\sigma_y+s\Delta\sigma_z$, where $s=\pm1$, $v_F$ is  the Fermi velocity, and $\Delta$ denotes  the energy gap. We assume that the electric and magnetic fields are polarized along the $y$- and $z$-axes, respectively, and both propagate along the $x$-axis, which is also the direction of the current, as
illustrated in Fig.~\ref{figure.1}. This setup corresponds to the index assignment  $\mu=\eta=x$ and  $\nu=\gamma=y$. 
In this configuration, the energy difference between the lower and upper bands is $\varepsilon_{ab}=-2\sqrt{(k^2_x+k^2_y)v^2_F+s^2\Delta^2}\equiv-2d$. The quantum metric components are then obtained as~\cite{zhang2022revealing} $g_{xy}=-k_xk_yv^4_F/4d^4$, $g_{yy}=v^2_F(k^2_xv^2_F+s^2\Delta^2)/4d^4$.
At zero temperature,  with the Fermi level lying above the band gap, the Ohmic and Hall conductivities described by Eqs.~(\ref{r7}) and (\ref{r8})  can be evaluated via polar coordinate integration, yielding explicit expressions for each valley
\begin{eqnarray}\label{61}
\sigma^{(O,0)}_{xy}&=&5e^3\frac{v^2_F(\epsilon^2_F-\Delta^2)(\epsilon^2_F+3\Delta^2)}{64\pi \epsilon^6_F}B_z(\mathbf{q}_0,\omega_0),\\
\sigma^{(H,0)}_{xy}&=&-e^3\frac{v^2_F(\epsilon^2_F-\Delta^2)}{4\pi \epsilon^4_F}B_z(\mathbf{q}_0,\omega_0),
\end{eqnarray}
where   $\epsilon_F$ denotes   the Fermi energy. In contrast, when the Fermi level  falls below the gap, both conductivities reverse their signs, as illustrated  in  Fig.~\ref{figure.2}. This figure presents the variation of these conductivities with the dimensionless Fermi energy $\epsilon_F/\Delta$, demonstrating that both conductivities can be efficiently tuned by adjusting the Fermi energy $\epsilon_F$. 
In addition, the magnitude of this bilinear magnetoelectric response exhibits a quadratic dependence on the ratio $v_F/\Delta$. Consequently,  increasing $v_F/\Delta$ can significantly enhance the  response. To quantitatively estimate the magnitude of this response, we adopt  typical parameter values.  Taking the Fermi velocity $v_F=10^6$ m/s and band gap $\Delta=0.1$ eV, the Ohmic conductivity can reach $7 \times 10^{-3}e^2/2\pi \hbar$. When the band gap is reduced to $\Delta=0.02$ eV,  this Ohmic conductivity  increases to
$0.17 e^2/2\pi \hbar$. Here,   $e^2/2\pi \hbar$ represents the fundamental conductance quantum characteristic of the quantum Hall effect.  This  demonstrates   that such an Ohmic response  can become  large in systems with a  high Fermi velocity and a narrow band gap.

\section{Conclusion}
We systematically studied nonlinear Ohmic transport in second-harmonic optical responses and second-order bilinear magnetoelectric responses. Using the Matsubara Green's function formalism from condensed matter quantum field theory, we built a solid quantum field-theoretic framework for describing nonlinear Ohmic transport. Our analysis yields  four main results.

First, we proved  that, for both types of responses, the Ohmic conductivity vanishes at linear order in the relaxation time, leaving only the Hall conductivity. Second, the nonlinear optical Ohmic conductivity contains a nonlinear Drude-like part and an intrinsic term controlled by the  fully symmetrized normalized quantum metric dipole, which exhibits a transverse character. Third, and more remarkably, we predicted a novel intrinsic Ohmic contribution as well as a pure Hall contribution arising purely from band geometry in the bilinear magnetoelectric response. This contribution, which can also be transverse in nature, takes an analytical form similar to but different from its optical counterpart. Fourth, using a two-dimensional Dirac model, we  demonstrated that this geometry-induced nonlinear Ohmic conductivity can be  sufficiently large  to be detected in materials with high Fermi velocity and narrow band gaps.

The finding of band-geometry-derived intrinsic Ohmic conductivity in bilinear magnetoelectric systems deepens our understanding of quantum geometric effects in quantum transport and offers a new way to probe  normalized quantum metric dipole in condensed matter. Possible extensions include spin systems~\cite{sinova2004universal,zhang2022geometric}, the skew scattering and side-jump mechanisms~\cite{2022Graphene,PhysRevLett.132.046303}, third-order response~\cite{mikhailov2016quantum}, and many-body effects  beyond the single-particle picture.~\cite{mahan2013many}.

\section*{Acknowledgements}
A. Z. acknowledges the support from Shanghai Magnolia Talent Plan Youth Project and Shanghai Normal University for initial start-up funding  (Grant No. 307-AF0102-26-005306). C. M. W was supported by the National Natural Science Foundation of China (Grant No. 12474048).

\section*{Data availability}
The data supporting this study’s findings are available
within the article.

\bibliographystyle{unsrt}
\bibliography{ref}

\end{document}